# Multiple soliton compression stages from competing plasma nonlinearities in mid-IR gas-filled hollow-core fibers

MD. SELIM HABIB*, CHRISTOS MARKOS, OLE BANG, AND MORTEN BACHE

*DTU Fotonik, Department of Photonics Engineering, Technical University of Denmark, Kgs. Lyngby, DK-2800*
*Corresponding author: seha@fotonik.dtu.dk*

**We investigate numerically soliton-plasma interaction in a noble-gas-filled silica hollow-core anti-resonant fiber pumped in the mid-IR at 3.0 μm. We observe multiple soliton self-compression stages due to distinct stages where either the self-focusing or the self-defocusing nonlinearity dominates. Specifically, the parameters may be tuned so the competing plasma self-defocusing nonlinearity only dominates over the Kerr self-focusing nonlinearity around the soliton self-compression stage, where the increasing peak intensity on the leading pulse edge initiates a competing self-defocusing plasma nonlinearity acting nonlocally on the trailing edge, effectively preventing soliton-formation there. As the plasma switches off after the self-compression stage, self-focusing dominates again, initiating another soliton self-compression stage in the trailing edge. This process is accompanied by supercontinuum generation spanning 1-4 μm. The technique could be exploited to generate an ultrafast sequence of several few-cycle pulses.** © 2016 Optical Society of America

http://dx.doi.org/10.1364/OL.99.099999

---

A new paradigm in nonlinear optics has emerged in hollow-core (HC) gas filled fibers [1], where the excitation of few- or even single-cycle temporal solitons with extreme peak intensities generate a plasma in the gas that affects the nonlinear dynamics *nonlocally*. Two competing nonlinearities are behind these phenomena as the temporal soliton relies on self-focusing self-phase modulation (SPM) effects (i.e. the nonlinear refractive index $n_2 > 0$), while the plasma generates a self-defocusing nonlinearity ($n_2 < 0$). The nonlocal nature of the self-defocusing nonlinearity occurs because it is the leading edge of the intense soliton that generates a plasma that affects the trailing edge of the soliton. This uniquely happens in a fiber geometry, where the soliton transverse mode is effectively described by the fiber mode, which allows long interaction lengths. The physics and nonlinear dynamics are therefore radically different from, e.g., a free-space focusing geometry generating a filament where the plasma and SPM also interact dynamically [2].

The light confinement inside hollow-core (HC) fibers filled with Raman inactive noble gases constitutes an efficient route to study interesting soliton-plasma dynamics. Initially gas-filled kagomé HC fibers [3–5] pumped in the near-IR were used as these were the first HC fibers to allow octave-spanning bandwidths as required to observe few-cycle solitons. However, only limited research has been carried out in the mid-IR range [6]. This range is interesting because the photon energy is much lower, promising different plasma formation dynamics, but this has yet to be investigated. Moreover, nonlinear optics in the mid-IR is currently a very active research field for supercontinuum generation [7,8] and few-cycle pulses [9], in particular exploiting filamentation in bulk media. The pulse energy limit of ~1 μJ of filamentation [8,9] was recently overcome by exploiting effective self-defocusing effects in bulk crystals [10]. HC gas-filled fibers insetad provide an interesting alternative as they also can sustain 10s of μJ pulse energies, tolerate muptiple-Watts of average powers, provide a clean spatial mode profile and give flexible beam handling and delivery.

Here we investigate the soliton-plasma dynamics in a mid-IR pumped HC fiber. Interestingly, we find a novel soliton dynamics scenario where multiple soliton self-compression stages are observed, which we explain as a direct consequence having distinct propagation stages with either the self-focusing or the self-defocusing nonlinearity dominates. This happens for certain system parameter ranges, mainly involving moderate gas pressure and input pulse intensities.

We use an HC fiber based on the so-called anti-resonant (AR) effect. HC-AR fibers provide relatively low-loss transmission, low light-glass overlap, and broadband guidance [11–14]. One of the main striking features of HC-AR fibers is that ~99.99% light can be guided inside the central hollow-core region, which significantly enhances the damage threshold levels [4,14]. Another advantage of using gas-filled HC fibers is that both the dispersion and nonlinearity can be tuned by simply changing the pressure of the gas [4,15,16] while at the same time providing extremely wide transparency ranges. Recently, silica HC-AR fibers suitable for the mid-IR were demonstrated with a propagation loss of <0.1 dB/m in the wavelength range 3-4 µm [11,12].

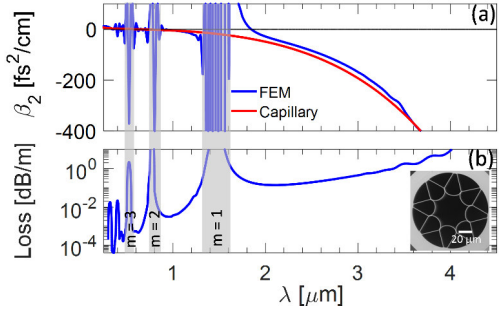

**Fig. 1.** Calculation of an HC-AR fiber filled with 1.2 bar Xe of (a) GVD vs. wavelength using MS capillary model (red curve) and FEM (blue curve), (b) propagation loss (using FEM). The gray bars indicate high loss regions. The inset shows an SEM of the HC-AR fiber used in our calculations, having 43.7 µm core diameter and 740 nm wall thickness.

Here we focus on the mid-IR properties of silica HC-AR fibers filled with the noble gas xenon (Xe). We pump at 3.0 µm and study the nonlocal soliton-plasma dynamics in a recently fabricated HC-AR fiber (see Fig. 1 (b) inset) having 43.7 µm core diameter and 0.74 µm silica wall thickness. This wall thickness introduces the first high-loss resonance band around 1.55 µm, and was chosen to give minimal loss at the 3.0 µm pump wavelength.

We calculated the group-velocity dispersion (GVD) using the Marcatili and Schmeltzer's (MS) capillary model [15,17], considering only the fundamental mode and ignoring polarization effects (similar to recent works [3,4,15]). The dispersion and leakage loss were calculated using the finite-element method (FEM). To accurately model the leakage loss, the mesh size and parameters for the perfectly-matched layer were carefully optimized. A maximum mesh size of λ/6 and λ/4 was used in silica and air regions, respectively [13,18]. The power overlap in the silica walls was used to estimate the effective material loss and then added to the leakage loss to obtain the final propagation loss [13].

Figure 1(a) shows the GVD of the HC-AR fiber filled with Xe at 1.2 bar pressure, calculated using the MS capillary model (red line) and FEM (blue line), while Fig. 1(b) shows the FEM calculated loss spectra. The FEM calculations are able to track the loss and the rapidly oscillating GVD in the high-loss resonance bands (gray shaded regions, with $m$ being the resonance band index), which are introduced by the resonant coupling between the core and cladding modes. In the resonance bands the GVD found by the MS capillary model stays smooth and continuous, while outside the bands it follows the GVD found by FEM quite well. The FEM results predict a propagation loss as low as ~ 0.45 dB/m at 3.0 µm.

The optical pulse propagation in the gas-filled HC-AR fiber was studied using the unidirectional pulse propagation equation [4,19]

$$\partial_z E = i(\beta(\omega) - \omega/v_g)E(z,\omega) - \alpha(\omega)/2$$
$$+ i\omega^2/(2c^2\varepsilon_0\beta(\omega))F[P_{NL}(z,t)] \quad (1)$$

where $z$ is the propagation direction, $t$ the time in the reference frame moving with the pump group velocity $v_g$, $E(z,\omega)$ the electric field in the frequency domain, $\omega$ the angular frequency, $\alpha(\omega)$ the propagation loss, $c$ the vacuum speed of light, $\beta(\omega)$ the propagation constant, $F$ denotes the Fourier transform. $P_{NL}(z,t)$ is the nonlinear polarization given by [4] $P_{NL}(z,t) = \varepsilon_0\chi^{(3)}E^3 + P_{ion}(z,t)$; the first term describes the Kerr effect, where $\varepsilon_0$ and $\chi^{(3)}$ are the vacuum permittivity and third-order nonlinear susceptibility, respectively, while the second term is the ionization effect expressed as [4]

$$P_{ion}(z,t) = \int_{-\infty}^{t}\frac{\partial N_e}{\partial t'}\frac{I_p}{E(z,t')}dt' + \frac{e^2}{m_e}\int_{-\infty}^{t}\int_{-\infty}^{t'}N_e(z,t^*)E(z,t'')dt''dt',$$
(2)

where $N_e$ is the free electron density, $m_e$ and $e$ are the mass and charge of an electron, and $I_p$ is the ionization energy of the gas. The nonlinear refractive index ($n_2$) of Xe at 1.2 bar was taken to be $8\times10^{-23}$ m$^2$/W and can as a good approximation be considered wavelength independent [20]. The third-order susceptibility was calculated using $\chi^{(3)} = (4/3)c\varepsilon_0 n_2 n^2$. The Raman contribution of silica was considered negligible due to the very low light-glass overlap (<<1%). The dynamics of $N_e(z,t)$ depends on whether multi-photon or tunnel ionization is occurring. In our case, the peak intensity inside the HC-AR fiber reaches around 70 TW/cm$^2$. This should be in the range where tunnel ionization dominates over multi-photon ionization, in particular because in the mid-IR the photon energy is low and thus multi-photon ionization has a high intensity onset threshold. Therefore, to calculate the free electron density, it is enough to consider quasi-static tunneling ionization based on the Ammosov, Delone, and Krainov (ADK) model. We also used the Perelomov, Popov, and Terent'ev (PPT) [4] model, which includes multi-photon ionization as well, and found essentially the same results. Finally the Keldysh parameter was checked to be less than unity. This conforms that tunnel ionization dominates.

The optical pulse propagation was modelled using either the full FEM loss and GVD profile (see appendix Fig. S1) or using no loss and the GVD from the MS capillary model (see Fig. 2). In the results presented below we chose to use the MS capillary model as it allows to better understand the fundamental dynamics behind the observed multi-compression stages without the interference oscillations due to the resonances of the FEM model. We emphasize that the two models qualitatively give similar results, as evidenced in the direct comparison between the two models (see appendix Fig. S1). Figure 2 shows the spectral and temporal soliton-plasma dynamics in a 25 cm HC-AR fiber filled with 1.2 bar Xe pumped in the anomalous dispersion regime at 3.0 µm with 100 fs, 20 µJ Gaussian pulses (typical parameters from emerging mid-IR optical parametric chirped pulse amplification laser systems). The low pressure was chosen to have a zero-dispersion wavelength (ZDW) at 630 nm deep into the visible, as the plasma dynamics tended to blue-shift the soliton dramatically well into the short-wavelength near-IR. Initially, the pulse propagation is dominated by the interplay between anomalous dispersion and self-focusing SPM, leading to strong soliton self-compression down to sub-single cycle duration of 7 fs (less than single-cycle duration at its blue-shifted center wavelength of ~1.5 µm) after 4.8 cm. It can be seen from Fig. 2(a) that at the maximum temporal compression point a blue-shifted spectrum is found, essentially forming a supercontinuum with a multiple octave-spanning bandwidth from 1.0-4.0 μm. The spectrum broadens mainly towards the blue due to plasma formation in the self-compression stage; this is known to blue-shift the soliton [5]. The plasma forms as the leading pulse field strength rises during the self-compression stage: The free electron density $N_e \sim 10^{23}$ m$^{-3}$ as shown in Fig. 2(c) is large enough to ionize the gas and form a plasma, and this renders the average nonlinear index change across the pulse negative, see the red curve in Fig. 2(c), while it remains positive before and after the soliton self compresses. The supercontinuum has also content below the ZDW consisting of dispersive waves (DWs) phase-matched to the soliton. After the maximum compression point, both the intensity and generation of free electrons drop below the ionization threshold level as shown in Fig. 2(c).

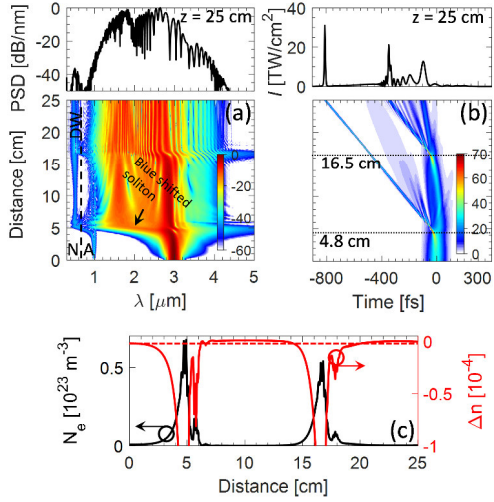

**Fig. 2.** Simulation showing (a) normalized power spectral density (PSD) [dB], (b) temporal intensity profile [TW/cm$^2$] and (c) average free electron density (left) and average nonlinear refractive index change (right) of a 20 μJ, 100 fs long pulse in a 43.7 μm core Xe-filled HC-AR fiber under 1.2 bar. The vertical dashed line in (a) indicates the location of ZDW = 630 nm) N: normal dispersion regime; A: anomalous dispersion regime, DW: Dispersive wave. The dashed red line in (c) indicates $\Delta n$=0. The simulation uses the MS capillary model, neglecting both resonances in the dispersion and linear propagation loss.

Interestingly, we then observe another soliton self-compression stage at around 16.5 cm. Let us describe the main mechanisms responsible for this. In Fig. 3 we plot intensity vs. time, the spectrogram and the normalized PSD vs. frequency at selected distances. Initially, Fig. 3(a) shows that at $z$=3 cm self-focusing SPM dominates giving a positive nonlinear chirp across the pulse (indicated by the dashed line through pulse center; in the wavelength vs. time spectrogram we remind that a negative slope correponds to a positive chirp in this representaion). At the maximum compression point ($z$=4.8 cm, see Fig. 3(b)), the pulse is compressed down to 7 fs. The nonlinear refractive index change ($\Delta n$, see the black curve inside the spectrogram) is here strongly negative across the trailing edge of the pulse due to the high intensity in the leading edge of the compressed pulse; this is the nonlocal action of the competing plasma-induced self-defocusing nonlinearity, which results in a negative chirp across the trailing pulse edge that prevents soliton compression of this part of the pulse. In contrast, at the early SPM stage at $z$=3 cm $\Delta n$ has only a weak negative value at the trailing edge. It should be emphasized that the maximum nonlinear refractive index change $\Delta n \approx -5 \times 10^{-4}$ and the maximum nonlinear plasma index is ~ 9 times higher than the nonlinear Kerr refractive index. Such a significant change in the nonlinear refractive index indicates that the plasma contribution is dominant over the Kerr effect. The nonlinear refractive index change between Kerr and plasma was calculated using [5]

$$\Delta n = n_2 I - \omega_p^2 / 2 n_0 \omega_0^2 . \qquad (3)$$

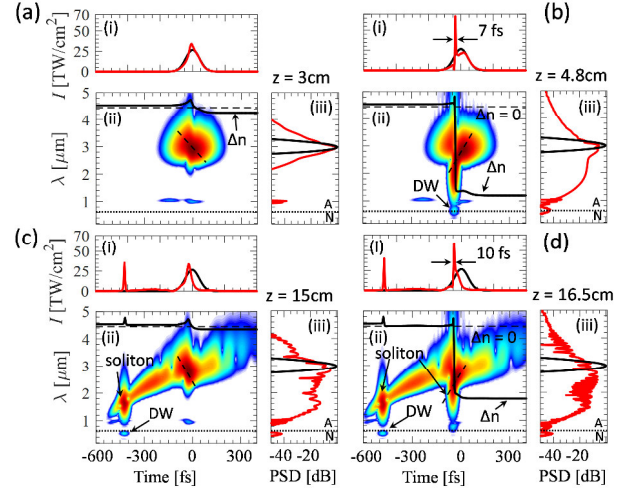

**Fig. 3.** Intensity vs. time (i), spectrogram (ii), and norm. PSD vs. frequency (iii) for (a) $z$=3 cm, (b) $z$=4.8 cm, (c) $z$=15 cm, and (d) $z$=16.5 cm of the simulation in Fig. 2. In the spectrogram the horizontal dashed and dotted black lines indicates $\Delta n$=0 and the ZDW, respectively, the slanted dashed black line through the pulse center indicates the slope of the chirp across the compressed pulse, and the black curve shows the nonlinear refractive index change. The spectrogram was calculated using a 25 fs Gaussian gate pulse. DW: dispersive wave. The spectrogram is represented by a dB scale from 0 dB to -50 dB.

The first term in Eq. (3) stems from the self-focusing Kerr effect, where $n_2$>0 is the nonlinear refractive index and $I$ is the pulse intensity. The second term is due to the plasma formation, where $n_0$ is the linear refractive index of the filling gas, $\omega_0$ is the central angular frequency, and $\omega_p$ is the plasma frequency expressed as [5] $\omega_p^2 = N_e e^2 / m_e \varepsilon_0$ where it is important to note that the free-electron density $N_e$ changes dynamically across the pulse. After the first compression, the soliton relaxes, leading to a drop in peak intensity (see Fig. 2(c)), and the free electron density drops quickly making the plasma disappear a few cm after the self-compression point. Essentially now a second stage starts with self-focusing SPM dominating (from $z$=6-15 cm). In Fig. 3(c) the spectrogram at $z$=15 cm shows that this allows the trailing edge to accumulate enough positive nonlinear phase shift to flip the chirp from negative to positive. This chirp is then subsequently compensated by the anomalous GVD to give another soliton self-compression stage at 16.5 cm, Fig. 3(d), this time down to 10 fs and 60 TW/cm$^2$ peak intensity. The same dynamics is now seen as in the first stage: a plasma forms during the compression that nonlocally induces a negative chirp on the trailing edge (dashed line), again preventing complete compression. The energy in the trailing edge is reduced compared to the first compression.

It is clear that the steep onset and extinction of the plasma around the soliton self-compression point and the nonlocal action of the plasma nonlinearity are key in explaining the multiple compression stages. During the first compression stage self-focusing SPM dominates, but as the plasma turns on ignited by the increasing intensity of the leading edge it induces in a nonlocal fashion a large negative chirp across the trailing edge, which prevents the soliton in forming symmetrically across the pump pulse profile. Consequently only little energy is then retained in the soliton. When the negative nonlinearity of the plasma is then subsequently turned off as the peak intensity of the soliton drops after the self-compression point, the peak power in the uncompressed trailing edge is large enough to initiate an SPM-induced self-focusing chirp-reversal stage leading to an overall positive chirp again and thus builds up to another soliton self-compression stage. This will repeat until the negatively-chirped trailing pulse edge has insufficient peak power to sustain the SPM chirp-reversal stage following the extinction of the self-defocusing plasma nonlinearity after the self-compression point; in this way one could engineer an ultrafast sequence of few-cycle pulses.

This phenomenon of multiple soliton self-compression stages is different from the pulse splitting observed in the near-IR [21]. The key is to ensure excitation of the plasma only around the soliton-self compression stage, giving distinct propagation stages where either the self-focusing or the self-defocusing nonlinearity dominates. This seems to happen when the gas pressure and pulse energy are not too high. In contrast, in the near-IR case [21] the self-defocsuing plasma nonlinearity always dominates on average, leading to pulse splitting at the first compression stage. However, the multiple soliton self-compression effect can also be observed in the near-IR by properly choosing the system parameters (see appendix Fig. S2). Similarly, pulse splitting can also happen in the mid-IR for high energies and pressure (see appendix Fig. S3).

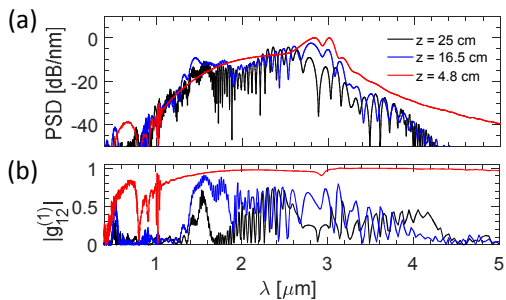

**Fig. 4.** (a) Power spectral density and (b) complex degree of first order coherence at z = 4.8 cm (red curve), z = 16.5 cm (blue curve), and z = 25 cm (black curve) obtain with 20 μJ, 100 fs long pulse filled Xe-filled HC-AR fiber under 1.2 bar. The averaged spectra and coherence properties were found by averaging over 20 simulations with random one photon per mode noise seeds.

The calculated coherence of the supercontinuum is shown in Fig. 4, visualized at suitable propagation distances. The coherence at the first compression stage for $z$=4.8 cm is high ($\approx$ 1 over a broad range), while the coherence drops for the second compression stage at $z$=16.5 cm and at the fiber end.

In conclusion, we presented a numerical investigation of the pulse propagation in a xenon-filled hollow-core anti-resonant silica fiber in the mid-IR in the high intensity regime. Due to the competing self-focusing and self-defocusing nonlinearities from the soliton-plasma interaction we found an intriguing multiple soliton self-compression stage dynamics. This was caused by a sudden onset and quenching of the plasma during and after self-compression allowing distinct propagation stages with either self-focusing or self-defocusing nonlinearity dominating, as well as the nonlocal action of the plasma, where the leading edge of the self-compressing self-focusing soliton ionizes the gas to affect the trailing edge with a competing self-defocusing nonlinearity. This process could be a novel way of generating a few-cycle pulse sequence. While we presented simulations using the simplified MS capillary model, we as mentioned found qualitatively similar results using a full FEM model, except for interference from phase-mathcing to narrow-band resonances. It is evident here there is a challenge in designing the anti-resonant wavelengths of the fiber to interfere as little as possible with the soliton-plasma dynamics.

**Acknowledgments**. We thank Dr. Rodrigo Amezuca-Correa and Dr. Jose Enrique Antonio-Lopez for providing the SEM image of HC-AR fiber. C.M. and O.B. acknowledge support from ShapeOCT (4107-00011A) and Danish Council for Independent Research (4184-00359B).

Appendix

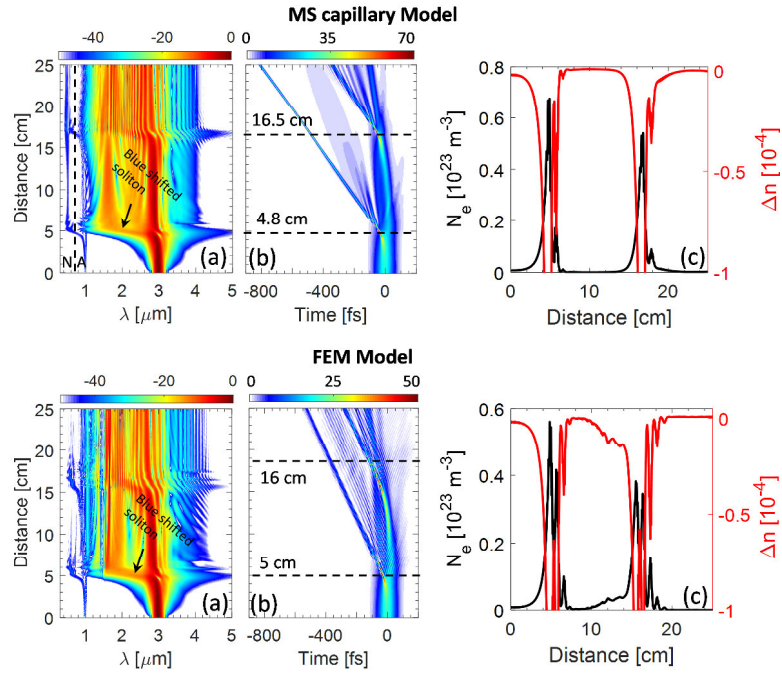

**Fig. S1.** Simulation showing (a) normalized power spectral density (PSD) [dB], (b) temporal intensity profile [TW/cm2], and (c) free electron density (left) and nonlinear refractive index change (right) as a function of propagation distance of a 20 µJ, 100 fs long pulse in a 43.7 µm core Xe-filled HC-AR fiber under 1.2 bar using MS capillary model (top) and FEM model (bottom). The vertical dashed line (top) indicates the location of the zero dispersion wavelength (ZDW = 0.63 µm). N: normal dispersion regime; A: anomalous dispersion regime.

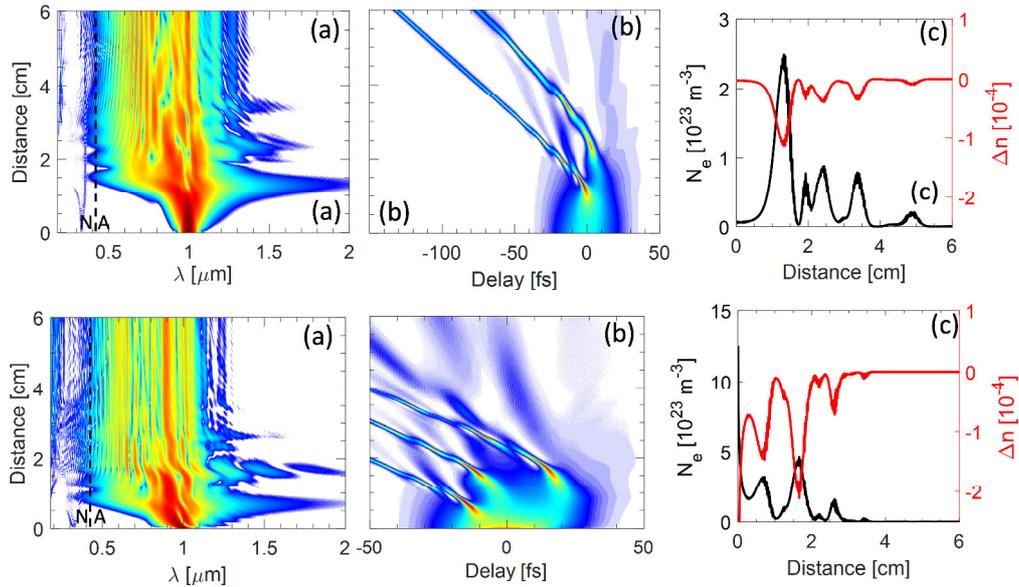

**Fig. S2.** (a) Spectral, (b) temporal evolution, and (c) free electron density (left) and nonlinear refractive index change (right) as a function of propagation distance of 33fs long pulse in a 14.56 µm core Xe-filled HC-AR fiber under 2 bar for 1 µJ (top) and 2 µJ (bottom). The vertical dashed line indicates the location of the zero dispersion wavelength (ZDW = 0.42 µm). N: normal dispersion regime; A: anomalous dispersion regime.

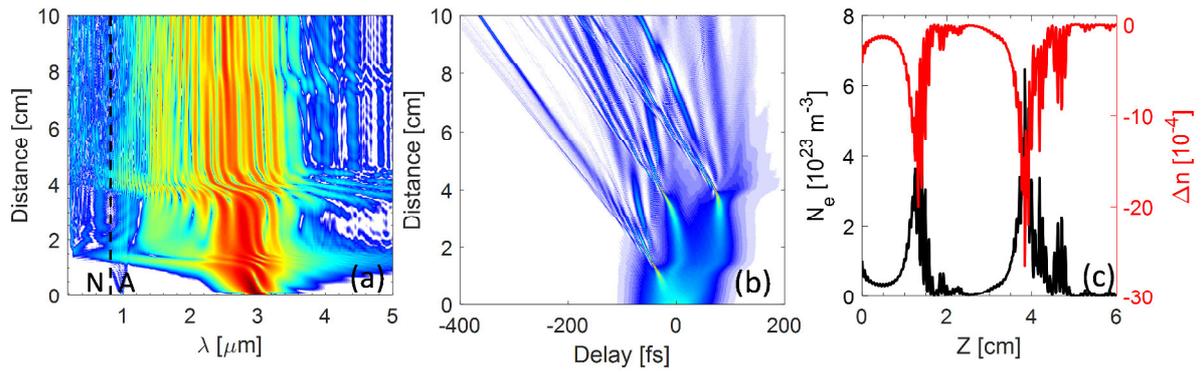

**Fig. S3.** (a) Spectral, (b) temporal evolution, and (c) free electron density (left) and nonlinear refractive index change (right) as a function of propagation distance of a 30 µJ, 100 fs long pulse in a 43.7 µm core Xe-filled HC-AR fiber under 3.5 bar. The vertical dashed line indicates the location of the zero dispersion wavelength (ZDW = 0.81 µm). N: normal dispersion regime; A: anomalous dispersion regime.